\documentclass[11pt]{article}

\usepackage[dvips]{graphicx}
\usepackage{amssymb,amsfonts,amsmath}

\usepackage{fullpage,amsthm,amsmath,times,color,authblk}
\usepackage[colorlinks=true,linkcolor=NavyBlue]{hyperref}
\usepackage{multibib}
\newcites{main}{References}

\newcommand{\commentidentity}[1]{}
\newcommand{\cut}[1]{{\color{Gray}{\commentidentity{#1}}}}

\usepackage[usenames,dvipsnames]{xcolor} 
\newcommand{\cdb}[1]{\commentidentity{({\color{WildStrawberry}CDB: {#1}})}}

\newcommand{\words}[1]{\commentidentity{({\color{CornflowerBlue} {#1} words})}}

\begin{document}

\title{Transdisciplinary electric power grid science\\{\large Opinion article + supporting information}}

%
%

\author[1,2]{Charles D.~Brummitt}
\author[3]{Paul D.~H.~Hines}
\author[4]{Ian Dobson}
\author[5]{Cristopher Moore}
\author[2,5,6]{Raissa M.~D'Souza}
\affil[1]{\small Department of Mathematics, University of California, Davis, CA 95616 USA}
\affil[2]{Complexity Sciences Center, University of California, Davis, CA 95616 USA}
\affil[3]{School of Engineering, University of Vermont, Burlington, VT 05405 USA}
\affil[4]{ECpE Department, Iowa State University, Ames IA 50011 USA}
\affil[5]{Santa Fe Institute, 1399 Hyde Park Road, Santa Fe, NM 87501 USA}
\affil[6]{Departments of Mechanical and Aerospace Engineering and of Computer Science, University of California, Davis, CA 95616 USA}


%
%
%
%
%


\date{\empty}


\maketitle

\begin{center}
\vspace{-.8cm}
{\large Pages 1--2 are the opinion article published in \href{http://www.pnas.org/content/110/30/12159}{PNAS {\bf 110} (30) 12159 (2013)}. \\ \vspace{2mm}
Pages 4--10 are supplementary material, with more explanation of the main components and feedbacks in electric power systems, as well as a list of useful sources of data on power grids.}\\
\end{center}

\setcounter{tocdepth}{1}
\tableofcontents 

\newpage 
\part{Opinion article}

When a tenth of humanity lost power over two days in India in July 2012, technical failure was not the only culprit. 
Like many recent blackouts, 
this outage resulted from couplings among systems, including extreme weather exacerbated by climate change, human operator errors, suboptimal policies, and market forces. Predictions that climate change intensifies droughts and tropical cyclones presage more weather-induced blackouts. 
Even without weather disasters, small disturbances can trigger cascading failures. 
Ill-designed electricity markets can cause blackouts~\citemain{Blumsack2010main}. So can dependence on cyber infrastructure.
\words{104}

Reliable electricity provides more than convenience; it fuels economies, governments, health care, education and poverty reduction. As populations shift to cities and consume more energy, confronting the multifaceted challenges to reliable electricity becomes paramount.
\words{34}

But enhancing reliability is no easy task. Upgrades can provide safety buffers, yet economic pressures can quickly consume new capacity. The opposing objectives of reliability and cost balance such that massive blackouts, albeit rare, continue to plague electric power systems~\citemain{Dobson2007main}. Regulation is not a panacea either: suppressing small blackouts, for instance, may increase the risk of large ones~\citemain{Dobson2007main}, a phenomenon that also plagues forest fires~\citemain{Brummitt2012main}. 
Because 
blackouts are inevitable, ensuring that critical services survive during blackouts should compete for investment with preventing them. 
Scientific understanding of decades-long feedbacks
should inform wise investment in robustness or resilience (e.g., harden against storm damage or build distributed generation?). 
\words{107}

The ``smart grid'', which monitors and controls electrical infrastructure in detail, promises enhanced reliability and efficiency, but it introduces concerns over privacy and cyber--physical interdependence. Furthermore, no computer can reliably control power grids that span continents. 
Thus, human operators also control the grid, occasionally making errors that trigger blackouts. 
We sorely lack convincing scientific analyses of the interactions among human operators, protocols, automatic controls and physical grids.
\words{68}

A third objective increasingly shapes power systems: sustainability. Renewable, distributed energy sources may confer resilience and mitigate climate change. But wind and solar's intermittent generation remains costly, and polluting fossil fuels fill the gap when renewables fall short of demand. 
Furthermore, moving power from windy or sunny locations to cities couples distant regions.
Connections among regions of a power grid spread risk, like in other infrastructures (e.g., default among banks or viruses among computers).
%
Using models to determine optimal connectivity is a problem that transcends disciplines~\citemain{Brummitt2012main}. 
\words{90} 

\cdb{number of words so far: 478 (not including figure caption nor section titles nor the references).}

 {\bf Tradeoffs of model complexity}:\ 
Power grid modeling is not merely an academic exercise; in fact, models keep the lights on. Every hour, operators run thousands of \cut{grid} simulations to determine the consequences of plausible disturbances. However, power grid failures are difficult to model because they involve so many complicated mechanisms. 
\words{46}

The tradeoffs of model complexity apply to any complex system\cut{, engineered or natural}. 
Stylized models can reveal the big picture and generate hypotheses, but throwing away rich data can relegate these models to irrelevance. For instance, many stylized models of cascading failures in power grids treat blackouts as epidemics that spread between adjacent nodes\cut{ (i.e., buses in a power grid)}. By contrast, real blackouts also spread non-locally: after a component fails, voltages and currents change throughout the network, potentially triggering failures hundreds of miles away. Thus, na\"ive, topological models of power grids offer little insight\cut{; the physics of electric power flow are essential}~\citemain{Hines2010main}.
\words{103}

At the other extreme, 
models resembling flight simulators capture many details (usually of the physics, not of the human operators and traders). 
Yet detailed models can obscure the important aspects and can require parameters difficult to measure. 
\words{55}

\cdb{words so far (not including figure caption): 682}

 {\bf Grand challenges}:\ 
Modeling the feedbacks in Fig.~\ref{fig:web} 
with accuracy but also insight 
is the first grand challenge for transdisciplinary power grid science. New smart grid measurement devices provide orders-of-magnitude more data with which to validate models. However, limited access to data severely impedes research. 
\words{43}

Validated models enable the next grand challenge: improve and transform power grids to meet 21$^\text{st}$-century pressures. Reliable electricity must reach more people demanding more energy in more places. 
Even in developed countries, reliability continues to lag. Robustness against threats from interdependence \cut{(Fig.~\ref{fig:web})} and \cut{from} malicious attacks will require transdisciplinary understanding. These challenges span engineering, physics, complex networks, computational science, economics, and social sciences.  
\words{65}
 
Ecology can also contribute. Like ecosystems, power grids consist of many species---generators, consumers, operators, traders---\cut{clumped in regions, }subject to environmental pressures\cut{ (Fig.\ \ref{fig:web})}. Multi-scale feedbacks drive both ecosystems and power grids to homeostasis or to collapse. \cut{It is surprising that 
systems so heterogeneous and so complex function as well as they do. }
\words{55}

Unlike biological systems, however, power grids cannot rely on natural selection to adapt\cut{ to the 21$^\text{st}$-century}. It is up to us. Scientists, engineers, policy makers and funding agencies must collaboratively tackle the challenges. 
Integrating and \cut{even} transcending disciplines will enable 
new ideas to 
shape the electric power system, the lifeblood of modern civilization. 

\bibliographystylemain{unsrt}

\newpage

\part{Supporting information}

\section{Illustration of the feedbacks}
\begin{figure*}[hbtp]
\centering
\vspace{1cm}
\includegraphics[width=.8 \textwidth]{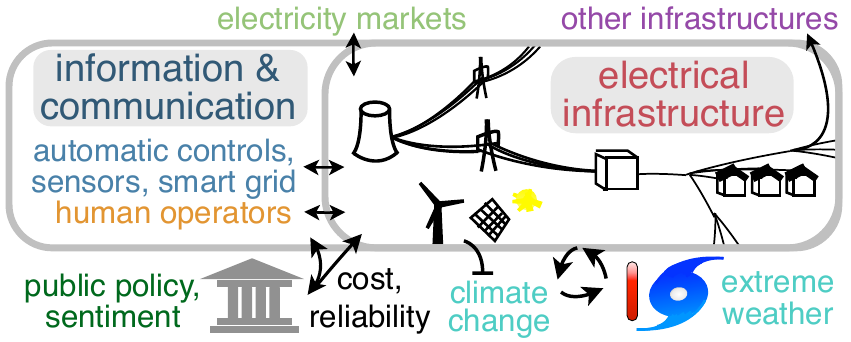}
\caption{Complex feedbacks surrounding electricity infrastructure.}
\label{fig:web}
\end{figure*}

Figure~\ref{fig:web} depicts the feedbacks surrounding electricity infrastructure, the component that engineers understand best. The web of interacts resembles an ecosystem. Human operators and automated systems control the grid to match generation with demand. Traders in markets bid on electric energy and trade electricity derivative contracts. Wind and solar generators are intermittent and typically far from dense populations, yet they mitigate the changes in climate that exacerbate blackout-causing extreme weather. Policy makers balance reliability, sustainability and cost as they invest in infrastructure, subsidize energy alternatives, and regulate operators and markets.

\section{Catastrophe from coupling}
\subsection{Recent systemic failures in power grids that stem from couplings among systems}
Two articles in IEEE Spectrum~\cite{Romero2012a,Romero2012b} highlight some causes of the blackout in India in July 2012. In particular, social and institutional failures---from imprudent policies for upgrading transmission lines to reluctance to disconnect overdrawing regions~\cite{Romero2012b}---played a key role. So did weather and perhaps climate change, because farmers pumped more water to alleviate drought~\cite{Romero2012b}.
Superstorm Sandy's costly consequences, including those from the electric power blackout in the northeastern United States, are summarized in Ref.~\cite{SandyEconomist}. 

For research articles that find that climate change is intensifying droughts, see Ref.~\cite{Hansen2012}, and for  tropical cyclones, see Refs.~\cite{Knutson2010,Grinsted2012}. The aforementioned blackouts in India and in the northeastern US are examples of blackouts partially caused by a drought and by a tropical cyclone, respectively.

Recent examples of cascading failures in developing countries include India 2012~\cite{Romero2012b} and Brazil 2012~\cite{Brazil2012}. For recent examples in developing countries, see 
Refs.\ \cite{2011SW_NERCFERC} (southwestern USA 2011),\ \cite{NERC2004} (northeastern North America 2003),\ \cite{UTCE:2004} (Italy 2003),\ \cite{UTCE:2007} (Europe 2006).

\subsection{Why confronting the challenges to reliable electric power is important}
Sixty percent of the global population is projected to live in urban areas in 2030~\cite{UNWorldPopulation2011}.
At the same time, global energy demand is projected to rise 50\% by 2030~\cite{GlobalTrends2030}. 

\section{More basics on the multilayered structure of electric power systems}\label{sec:basics}

Power grids comprise more than physical infrastructure, as illustrated in Fig.~\ref{fig:web}; they are also social and market systems, and challenges span all three of the physical, human and market layers. 
At the center lies what engineers understand best: the electrical infrastructure. Next, information and communication systems monitor and control the electrical infrastructure at increasing levels of detail using new technology. Novel capabilities of this cyber-physical ``smart grid'' promise to enhance reliability and efficiency, yet they risk introducing privacy concerns, security vulnerabilities, and dependence between electricity and information infrastructure. 

\subsection{How do human operators monitor and control power grids}  
In hundreds of control rooms across the United States, for instance, human operators augment automatic control systems by monitoring an array of computer screens and by controlling the grid to match generation with demand that varies throughout the day. The control centers coordinate with one another electronically and by telephone. Typically, automatic and manual controllers balance supply and demand and respond to contingencies without problems, but occasionally human controllers (or software or hardware) make errors that cause blackouts~\cite{NERC2004,2011SW_NERCFERC,SwissItalianBlackout}.

\subsection{Electricity markets} 
Humans also affect power grids through electricity markets---and in challenging new ways due to deregulation and due to the smart grid. Over the past few decades, power systems around the world have shifted to deregulated markets, in which suppliers compete by making bids to sell electric energy. But this competition has often fallen short on its promises of price reduction and better investment because market volatility makes capital expensive, investment needs quick return, and congestion becomes profitable~\cite{Blumsack2010}. In light of market catastrophes like the 2000--01 California energy crisis, the shift to deregulation continues but in measured ways. The smart grid will simultaneously complicate and streamline these markets by enabling a Cambrian explosion of new species, from small-scale generators that inject renewable energy into the grid, to smart appliances that bid for energy within price limits set by their owners. 

 \subsection{Electricity production's greenhouse gas emissions and climate change}
Feedback loops between electricity infrastructure and climate are slower but no less significant than the interactions with markets. Generating electricity produces more greenhouse gas emissions than any other economic sector (e.g., 34\% in the US~\cite{USgreenhousegas}, 26\% in the EU~\cite{EUgreenhousegas}, 24\% globally~\cite{UNEPClimateNeutrality}). These greenhouse gases contribute to global climate change, which poses significant risks to human health and welfare~\cite{IPCC2007_HumanHealth}. Furthermore, climate change jeopardizes reliable electric power because it intensifies the droughts~\cite{Hansen2012} and tropical cyclones~\cite{Knutson2010,Grinsted2012} 
that cause blackouts~\cite{Romero2012b,SandyEconomist}. 

\subsection{Grid reliability, the $n$$-$$1$ criterion, and similarity to systemic risk in other fields}
A key reliability standard intended to mitigate risk of cascading failures is the ``$n$$-$$1$ criterion'', which requires that blackouts not result from the failure of any single component in the grid. A major challenge is that hundreds of organizations operate portions of each synchronous, interconnected grid (of which there are three in the US), and no organization knows the state of the entire grid. Recent blackouts spreading from one area to another~\cite{Romero2012b,2011SW_NERCFERC} prompted discussion of expanding the $n$$-$$1$ criterion to require greater coordination among control areas. Designing and operating the interconnected control areas so that they collectively reduce the system-wide risk of cascading failure, without compromising their local performance, presents formidable challenges. Risk spreads in other infrastructures, too, such as default among banks or computer viruses among Autonomous Systems of the Internet. Thus, using models to determine optimal interdependence and control is a problem that transcends disciplines~\cite{Brummitt2012,Battiston2012}. 
\section{Tradeoffs of model complexity}\
Examples of the many complicated mechanisms involved in blackouts include thermal overloads, relay failure, voltage collapse, dynamic instability and operator error~\cite{Eppstein2012}. State-of-the-art models capture only a subset of these failure mechanisms; for examples, see Refs.~\cite{Eppstein2012,Nedic2006} and the review article on cascading failures in electric power systems~\cite{IEEECascadingFailureReview}. 

\section{Grand challenges for a transdisciplinary science of electric power systems}
\subsection{Starting points for finding data on power grids}
Though power grid data is difficult to obtain, there is some available, useful data:

\begin{itemize}
\item The freely available software MATPOWER~\cite{MATPOWER} provides a useful starting point; for instance, it contains data on the Polish transmission system, which has been used in academic papers (e.g.,~\cite{Eppstein2012})
\item Several of the US Independent System Operators (ISOs) provide information about their historical load, load forecast, and economic data. For example, see
\url{http://www.iso-ne.com/markets/hrly_data/}.
\item The National Renewable Energy Laboratory (NREL) recently published large sets of simulated wind speed and power data. See, for example,
\url{http://www.nrel.gov/electricity/transmission/eastern_wind_methodology.html}.
\item The Bonneville Power Administration (BPA) publishes some good data about transmission outages: 
\url{http://transmission.bpa.gov/business/operations/Outages/default.aspx}

\item In an effort to address the challenge of obtaining data for research and industry analysis, the Energy Information Administration (EIA) is proposing new data collection activities that will begin in 2014; for information on this effort see 
\url{http://www.eia.gov/survey/changes/electricity/}
\end{itemize}


\subsection{Examples of reliable electricity needing to reach more people}
A prominent example of the need to provide more electricity (and with greater reliability) is India, where 300 million people lack electricity~\cite{Romero2012b,GlobalTrends2030}. 
Developed countries continue to suffer outages, too; in the US, for instance, about a million people lose power every day on average~\cite{Eto2012}. 

\subsection{Threats from terrorist attacks that leverage interdependence among components of the electric power system}
The National Academy of Sciences report on robustness and resilience of the electric power system in the United States~\cite{NASreport} highlights dangers from the power system's age, inadequate guards against malicious attack, and interdependence with other infrastructure (like wireless communication), all of which exacerbate the risks of costly blackouts caused by extreme weather~\cite{SandyEconomist} or by terrorist attack~\cite{NASreport}. 
Whereas blackouts lasting days typically cost millions of dollars, blackouts lasting months (because of damage to hard-to-replace transformers, for example) could cost billions~\cite{NASreport,ForeignPolicy_ForgetRevolution}.

\section{Why ``transdisciplinary''?}
\begin{figure}[hbtp]
\begin{center}
\includegraphics[width=.9 \textwidth]{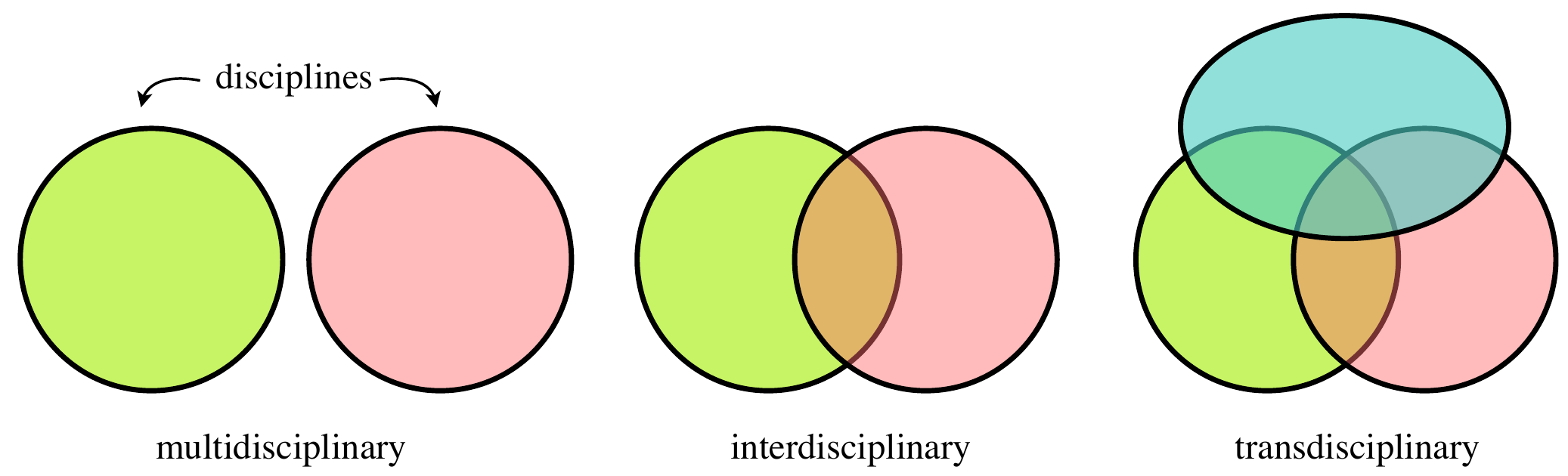}
\caption{
One way to illustrate the difference between multidisciplinary, interdisciplinary and transdisciplinary uses Venn diagrams~\cite{Choi2006}. With shaded circles denoting a discipline (such as ``physics''), multidisciplinary problems subdivide into independent problems that fall into different disciplines, whereas interdisciplinary problems require blending and creating new knowledge between fields (such as bioinformatics, quantum computing and astrobiology). Transdisciplinary research also blends disciplines but creates new knowledge outside of them.}
\label{fig:discipline}
\end{center}
\end{figure}

A useful resource for 
the definitions of multidisciplinary, interdisciplinary and transdisciplinary is Choi et al., 2006~\cite{Choi2006} and two followup papers~\cite{Choi2007,Choi2008}. For an illustration of these terms, see Fig.~\ref{fig:discipline}. 

In short, multidisciplinary problems sub-divide into independent subproblems that can be solved by practitioners in different disciplines, without much need for communication between the disciplines. By contrast, interdisciplinary problems require blending multiple fields and hence the creation of new knowledge between two fields. Transdisciplinary problems also require blending fields, but they create entirely new knowledge and techniques outside those fields. Tackling problems with large-scale feedbacks among different systems and with commonalities among other complex systems often becomes transdisciplinary. 
When it is not clear which kind of approach applies, Choi et al.~\cite{Choi2006,Choi2007,Choi2008} recommend using the word ``multiple disciplinary''.  

Some of the challenges highlighted in the opinion article may be multidisciplinary; for example, research into mitigating climate change and into reducing emissions from electricity generation can proceed rather independently (multidisciplinary). Other problems will be interdisciplinary. The smart grid, for instance, combines communication and electrical infrastructure together with new market capabilities (see Sec.~\ref{sec:basics}), so understanding these systems will likely blend these disciplines. For example, monitoring smart grids may benefit from knowledge about percolation in random graphs~\cite{Yang2012}. 
Models of feedbacks on long spatiotemporal scales, such as of reliability and cost~\cite{Dobson2007,Nedic2006}, and models that address problems with analogs in other complex systems, such as optimal connectivity~\cite{Brummitt2012}, begin to leave disciplinary boundaries (transdisciplinary). 


\section*{Acknowledgments}
We thank the attendees of the workshop Power Grids as Complex Networks~\cite{SFIWorkshop} for useful discussion that motivated this article, and we thank the Santa Fe Institute, Los Alamos National Laboratory, the Flora Foundation and the University of Vermont for funding the workshop. C.B.\ was partially supported by the Department of Defense (DoD) through the National Defense Science \& Engineering Graduate Fellowship (NDSEG) Program; P.H.\ by the US Department of Energy grant DE-OE0000447; I.D.\ by National Science Foundation grant CPS-1135825; C.M.\ by Air Force Office of Scientific Research and Defense Advanced Research Projects Agency grant FA9550-12-1-0432; R.D.\ by the Defense Threat Reduction Agency Basic Research Award HDTRA1-10-1-0088 and by the Army Research Laboratory under Cooperative Agreement Number W911NF-09-2-0053.

\bibliographystyle{unsrt}


\begin{thebibliography}{1}

\bibitem{Blumsack2010main}
Blumsack S (2010) {How the free market rocked the grid}. \emph{IEEE Spectrum}
  47:44--59.

\bibitem{Dobson2007main}
Dobson I, Carreras BA, Lynch VE, Newman DE (2007) Complex systems analysis of
  series of blackouts: Cascading failure, critical points, and
  self-organization. \emph{Chaos} 17.

\bibitem{Brummitt2012main}
Brummitt CD, D'Souza RM, Leicht EA (2012) {Suppressing cascades of load in
  interdependent networks.} \emph{Proc~Natl~Acad~Sci~USA} 109:E680--E689.

\bibitem{Hines2010main}
Hines P, Cotilla-Sanchez E, Blumsack S (2010) {Do topological models provide
  good information about electricity infrastructure vulnerability?}
  \emph{Chaos} 20.

\end{thebibliography}

\begin{thebibliography}{34}

\bibitem{Romero2012a}
Joshua~J. Romero.
\newblock {Planned Upgrade Set the Stage for Indian Blackout}.
\newblock {\em IEEE Spectrum}, August 2012.

\bibitem{Romero2012b}
J.~J. Romero.
\newblock {Blackouts illuminate India's power problems}.
\newblock {\em IEEE Spectrum}, 49(10):11--12, October 2012.

\bibitem{SandyEconomist}
R.~A.
\newblock Costs to come.
\newblock {\em The Economist}, October 2012.

\bibitem{Hansen2012}
James Hansen, Makiko Sato, and Reto Ruedy.
\newblock Perception of climate change.
\newblock {\em Proc.~Natl.~Acad.~Sci.~U.S.A.}, 109(37):E2415--E2423, 2012.

\bibitem{Knutson2010}
Thomas~R Knutson, John~L McBride, Johnny Chan, Kerry Emanuel, Greg Holland,
  Chris Landsea, Isaac Held, James~P Kossin, A~K Srivastava, and Masato Sugi.
\newblock {Tropical cyclones and climate change}.
\newblock {\em Nature Geoscience}, 3(3):157--163, 2010.

\bibitem{Grinsted2012}
Aslak Grinsted, John~C. Moore, and Svetlana Jevrejeva.
\newblock {Homogeneous record of Atlantic hurricane surge threat since 1923}.
\newblock {\em Proc.~Natl.~Acad.~Sci.~U.S.A.}, 109(48):19601--19605, 2012.

\bibitem{Brazil2012}
Tatiana Maria~T. de~S.~Alves.
\newblock Brazilian blackout 2012.
\newblock {\em Protection, Automation and Control World}, June 2012.

\bibitem{2011SW_NERCFERC}
Federal Energy Regulatory Commission.
\newblock {\em {Arizona-Southern California Outages on September 8, 2011}},
  April 2012.

\bibitem{NERC2004}
NERC.
\newblock {How and Why the Blackout Began in Ohio}.
\newblock In {\em Final Report on the August 14th Blackout in the United States
  and Canada}, chapter~5, pages 45--72. U.S.-Canada Power System Outage Task
  Force, April 2004.

\bibitem{UTCE:2004}
UTCE.
\newblock {Final Report of the Investigation Committee on the 28 September 2003
  Blackout in Italy}.
\newblock Technical report, Union for the Co-ordination of Transmission of
  Electricity, 2004.

\bibitem{UTCE:2007}
UTCE.
\newblock {Final Report: System Disturbance on 4 November 2006}.
\newblock Technical report, Union for the Co-ordination of Transmission of
  Electricity, 2007.

\bibitem{UNWorldPopulation2011}
{United Nations Department of Economic and Social Affairs/Population Division}.
\newblock {World Urbanization Prospects: The 2011 Revision}, 2011.
\newblock \url{http://esa.un.org/unup/pdf/WUP2011_Highlights.pdf}.

\bibitem{GlobalTrends2030}
{National Intelligence Council}.
\newblock {Global Trends 2030: Alternative Worlds}, December 2012.

\bibitem{SwissItalianBlackout}
{Commision de R\'egulation de L'\'Energie (CRE) and Autorit\`a per l'energia
  electrica e il gas (AEEG)}.
\newblock {Report on the events of September 28th culminating in the separation
  of the Italian power system from the other UCTE networks}.
\newblock Technical report, {CRE and AEEG}, April 2004.

\bibitem{Blumsack2010}
S~Blumsack.
\newblock {How the free market rocked the grid}.
\newblock {\em IEEE Spectrum}, 47:44--59, December 2010.

\bibitem{USgreenhousegas}
{United States Environmental Protection Agency}.
\newblock Sources of greenhouse gas emissions: Electricity sector emissions,
  June 2012.

\bibitem{EUgreenhousegas}
{European Environment Agency}.
\newblock {Why did greenhouse gas emissions increase in the EU in 2010? EEA
  analysis in brief}, May 2012.

\bibitem{UNEPClimateNeutrality}
GRID-Arendal.
\newblock {Kick the Habit: A UN Guide to Climate Neutrality}, 2008.
\newblock
  \url{http://www.grida.no/graphicslib/detail/world-greenhouse-gas-emissions-by-sector_6658}.

\bibitem{IPCC2007_HumanHealth}
U.~Confalonieri, B.~Menne, R.~Akhtar, K.L. Ebi, M.~Hauengue, R.S. Kovats,
  B.~Revich, and A.~Woodward.
\newblock {\em {Contribution of Working Group II to the Fourth Assessment
  Report of the Intergovernmental Panel on Climate Change}}, chapter~8, pages
  391--431.
\newblock Cambridge University Press, Cambridge, UK, 2007.

\bibitem{Brummitt2012}
Charles~D Brummitt, Raissa~M D'Souza, and E~A Leicht.
\newblock {Suppressing cascades of load in interdependent networks.}
\newblock {\em Proc.~Natl.~Acad.~Sci.~U.S.A.}, 109(12):E680--E689, February
  2012.

\bibitem{Battiston2012}
Stefano Battiston, Domenico~Delli Gatti, Mauro Gallegati, Bruce Greenwald, and
  Joseph~E Stiglitz.
\newblock {Liaisons dangereuses Increasing connectivity, risk sharing, and
  systemic risk}.
\newblock {\em Journal of Economic Dynamics and Control}, 36(8):1121--1141,
  August 2012.

\bibitem{Eppstein2012}
Margaret~J Eppstein and Paul D~H Hines.
\newblock {A ``Random Chemistry'' Algorithm for Identifying Collections of
  Multiple Contingencies That Initiate Cascading Failure}.
\newblock {\em Power Systems, IEEE Transactions on}, 27(3):1698--1705, 2012.

\bibitem{Nedic2006}
Dusko~P. Nedic, Ian Dobson, Daniel~S. Kirschen, Benjamin~A. Carreras, and
  Vickie~E. Lynch.
\newblock Criticality in a cascading failure blackout model.
\newblock {\em International Journal of Electrical Power \& Energy Systems},
  28(9):627--633, 2006.

\bibitem{IEEECascadingFailureReview}
R.~Baldick, B.~Chowdhury, I.~Dobson, Zhaoyang Dong, Bei Gou, D.~Hawkins,
  H.~Huang, M.~Joung, D.~Kirschen, Fangxing Li, Juan Li, Zuyi Li, Chen-Ching
  Liu, L.~Mili, S.~Miller, R.~Podmore, K.~Schneider, Kai Sun, D.~Wang, Zhigang
  Wu, Pei Zhang, Wenjie Zhang, and Xiaoping Zhang.
\newblock Initial review of methods for cascading failure analysis in electric
  power transmission systems ieee pes cams task force on understanding,
  prediction, mitigation and restoration of cascading failures.
\newblock In {\em Power and Energy Society General Meeting -- Conversion and
  Delivery of Electrical Energy in the 21st Century, 2008 IEEE}, pages 1--8,
  july 2008.

\bibitem{MATPOWER}
R.D. Zimmerman, C.E. Murillo-S\'anchez, and R.J. Thomas.
\newblock {MATPOWER: Steady-State Operations, Planning, and Analysis Tools for
  Power Systems Research and Education}.
\newblock {\em Power Systems, IEEE Transactions on}, 26(1):12--19, February
  2011.

\bibitem{Eto2012}
Joseph~H. Eto, Kristina~H. LaCommare, Peter Larsen, Annika Todd, and Emily
  Fisher.
\newblock Distribution-level electricity reliability: Temporal trends using
  statistical analysis.
\newblock {\em Energy Policy}, 49(0):243--252, 2012.
\newblock Special Section: Fuel Poverty Comes of Age: Commemorating 21 Years of
  Research and Policy.

\bibitem{NASreport}
{Committee on Enhancing the Robustness and Resilience of Future Electrical
  Transmission and Distribution in the United States to Terrorist Attack; Board
  on Energy and Environmental Systems; Division on Engineering and Physical
  Sciences; National Research C}.
\newblock {\em Terrorism and the Electric Power Delivery System}.
\newblock The National Academies Press, 2012.

\bibitem{ForeignPolicy_ForgetRevolution}
Douglas Birch.
\newblock {Forget Revolution}.
\newblock {\em Foreign Policy}, October 2012.

\bibitem{Choi2006}
B~C~K Choi and A~W~P Pak.
\newblock {Multidisciplinarity, interdisciplinarity and transdisciplinarity in
  health research, services, education and policy: 1. Definitions, objectives,
  and evidence of effectiveness.}
\newblock {\em Clinical and investigative medicine. Medecine clinique et
  experimentale}, 29(6):351, 2006.

\bibitem{Choi2007}
B~C~K Choi and A~W~P Pak.
\newblock {Multidisciplinarity, interdisciplinarity, and transdisciplinarity in
  health research, services, education and policy: 2. Promotors, barriers, and
  strategies of enhancement}.
\newblock {\em Clinical {\&} Investigative Medicine}, 30(6):E224--E232, 2007.

\bibitem{Choi2008}
B~C~K Choi and A~W~P Pak.
\newblock {Multidisciplinarity, interdisciplinarity, and transdisciplinarity in
  health research, services, education and policy: 3. Discipline,
  inter-discipline distance, and selection of discipline}.
\newblock {\em Clinical {\&} Investigative Medicine}, 31(1):E41--E48, 2008.

\bibitem{Yang2012}
Yang Yang, Jianhui Wang, and Adilson Motter.
\newblock {Network Observability Transitions}.
\newblock {\em Physical Review Letters}, 109(25):258701, December 2012.

\bibitem{Dobson2007}
I.~Dobson, B.~A. Carreras, V.~E. Lynch, and D.~E. Newman.
\newblock Complex systems analysis of series of blackouts: Cascading failure,
  critical points, and self-organization.
\newblock {\em Chaos}, 17(026103), 2007.

\bibitem{SFIWorkshop}
{Santa Fe Institute Workshop}.
\newblock Power grids as complex networks: Formulating problems for useful
  science and science-based engineering.
\newblock Santa Fe Institute, Santa Fe, NM, May 17--19, May 2012.
\newblock Funded by the Santa Fe Institute, Los Alamos National Laboratory, the
  University of Vermont, and the Flora Foundation.

\end{thebibliography}

\renewcommand{\refname}{Supporting Information References}

\end{document}